\documentclass[twocolumn,aps,superscriptaddress,showpacs,showkeys,prb]{revtex4-1}
\usepackage[utf8]{inputenc}
\usepackage{xspace}
\usepackage{color}
\usepackage{graphicx}
\usepackage{bm}
\usepackage{amsmath,amsfonts,amssymb}
\usepackage{upgreek}
\usepackage[colorlinks=true]{hyperref}
\usepackage{siunitx}
\DeclareSIUnit\muB{$\mu_B$}
\DeclareSIUnit\atper{{at.\,\percent}}
\newcommand{\sco}[1]{SrCoO$_{#1}$\xspace}

\newcommand{\eg}{\ensuremath{e_{\text{g}}}}
\newcommand{\tg}{\ensuremath{t_{2\text{g}}}}
\newcommand{\due}{\eg^\uparrow}
\newcommand{\dde}{\eg^\downarrow}
\newcommand{\dut}{\tg^\uparrow}

\newcommand{\dxy}{\ensuremath{d_{xy}}}

\newcommand{\duxy}{\ensuremath{d^{\uparrow}_{xy}}}

\newcommand{\ddxy}{\ensuremath{d^{\downarrow}_{xy}}}
\newcommand{\ddxz}{\ensuremath{d^{\downarrow}_{zx}}}
\newcommand{\ddyz}{\ensuremath{d^{\downarrow}_{yz}}}

\newcommand{\sikelvin}[1]{\SI{#1}{\kelvin}\xspace}
\newcommand{\sieV}[1]{\SI{#1}{\electronvolt}\xspace}
\newcommand{\simeV}[1]{\SI{#1}{\milli\electronvolt}\xspace}
\newcommand{\simB}[1]{\SI{#1}{\muB}\xspace}
\newcommand{\Ueff}{U_\text{eff}}
\newcommand{\EF}{E_\text{F}}

\newcommand{\Tc}{\widetilde{T}} 
\newcommand{\TC}{T_\mathrm{C}} 
\newcommand{\TN}{T_\mathrm{N}} 
\newcommand{\vo}{V$_\text{O}$\xspace}
\renewcommand{\vec}[1]{{{\bm #1}}}
\newcommand{\mypi}{\uppi}
\newcommand{\diff}{\text{d}}
\newcommand{\trace}{\text{Tr}}
\newcommand{\scatpaop}[1]{\hat{\tau}^{#1}}
\newcommand{\scatpaopup}[1]{\hat{\tau}^{#1}_{\uparrow}}
\newcommand{\scatpaopdw}[1]{\hat{\tau}^{#1}_{\downarrow}}
\newcommand{\singscatopup}[1]{\hat{t}^{-1}_{#1\uparrow}}
\newcommand{\singscatopdw}[1]{\hat{t}^{-1}_{#1\downarrow}}

\newcommand{\rspm}{RS$_\text{PM}$\xspace}
\newcommand{\rsfm}{RS$_\text{FM}$\xspace}
\newcommand{\mathcomma}{\,,}  		
\newcommand{\etal}{\textit{et al.}\xspace}

\newcommand{\add}[1]{{\color{black}{#1}}}

\newcommand{\affmluDE}{Institut f\"ur Physik, Martin-Luther-Universit\"at
  Halle-Wittenberg, Von-Seckendorff-Platz 1, 06120 Halle, Germany}

\newcommand{\affmpihalleDE}{Max-Planck-Institut f\"ur Mikrostrukturphysik,
  Weinberg 2, 06120 Halle, Germany}

\begin{document}
\hyphenation{an-ti-fer-ro-mag-ne-tic}
\title{Magnetic properties of defect-free and oxygen-deficient cubic SrCoO$_{3-\delta}$}
  \author{Martin Hoffmann}
  \email{mart.hoffi@gmail.com}
  \affiliation{\affmluDE}
  \affiliation{\affmpihalleDE}

  \author{Vladislav S. Borisov}
  \affiliation{\affmluDE}
  \affiliation{\affmpihalleDE}

  \author{Sergey Ostanin}
  \affiliation{\affmpihalleDE}

  \author{Ingrid Mertig}
  \affiliation{\affmluDE}
  \affiliation{\affmpihalleDE}
  
  \author{Wolfram Hergert}
  \affiliation{\affmluDE}

  \author{Arthur Ernst}
  \affiliation{\affmpihalleDE}

\date{\today}
 
\begin{abstract}
  We investigated theoretically electronic and magnetic properties of the perovskite
  material SrCoO$_{3-\delta}$ with $\delta\leq 0.15$ \add{using 
  a projector-augmented plane-wave method and a Green's function method.}
  This material is known from
  various experiments to be ferromagnetic with a Curie temperature of
  \SIrange{260}{305}{\kelvin} and a magnetic moment of
  \SIrange{1.5}{3.0}{\muB}.  Applying the magnetic force theorem as it
  is formulated within the Green's function method, we calculated 
  for SrCoO$_{3-\delta}$ the
  magnetic exchange interactions and estimated the Curie
  temperature. Including correlation effects by an effective $U$
  parameter within the GGA$+U$ approach \add{and verifying this 
  by hybrid functional calculations}, we obtained the Curie
  temperatures in dependence of the oxygen-deficiency close to the
  experimental values.
\end{abstract}
\maketitle
\section{Introduction}
Perovskite materials with the simple structure formula $AB$O$_3$
attracted attention in the last decades because of potential
applications in spintronics.  Especially, ferromagnetic metallic
perovskites have rekindled interest, since they can be used as
electrodes in complex oxide heterostructures, while conventional
ferromagnetic materials such as $3d$ transition metals are hardly
compatible with most of the oxides.

Over the last years, advanced experimental growing techniques allow
for the growth of heterostructures and multilayer \add{systems with a huge
variety of properties,} e.g. multiferroic,
magnetoelectric or magnetooptic.  Out of this class of materials, we
concentrated on \sco{3-\delta} \add{(SCO)} with a possible oxygen-deficiency
$\delta$.  In its pure single-crystal composition, the experiments
verify ferromagnetic behavior up to room temperature and metallic
conductance.\cite{Long2011jpcm} Metallic oxides are highly desired for
contacts and electrodes of \add{the above discussed} multi-functional heterostructures
because of a good lattice match between the electrode and the
top layer oxide. Undesired distortions of thin layers could be
reduced.  Simultaneously, it is possible to transport an electric
current, e.g. in the compounds SrCo$_{1-x}$Fe$_x$O$_3$ for $x\leq0.5$,
which are particularly suitable as electrode materials for solid
electrolyte oxygen sensors.  \cite{Shuk1993} In addition, complex
oxides based on \sco{3} like e.g.  La$_{0.6}$Sr$_{0.4}$CoO$_3$ or
SrCo$_{0.9}$Sb$_{0.1}$O$_{3-\delta}$ appear to be particularly useful
as cathodes for intermediate-to-low temperature solid oxide fuel
cells.\cite{Inagaki2000,Lin2009}

In the literature there are several experimental investigations and
theoretical studies on SCO. They all verify the ferromagnetic and
metallic behavior of this material.  In theory, in particular, the
electronic structure,\cite{MathiJaya1991prb,Zhuang1998,Wu2012} the spin
state,\cite{Potze1995,Kunes2012a} and possible lattice
distortions\cite{Lee2011} are \add{thoroughly} discussed.
Potze~\etal\cite{Potze1995} show that \sco{3} exhibits an intermediate
spin \add{(IS)} state due to the competition of intra-atomic exchange and the
cubic crystal field.  Such an IS state may be
understood as a high spin state of a Co$^{3+}$ ion
antiferromagnetically coupled to a ligand hole of $\eg$ symmetry
(notation from Ref.~\onlinecite{Potze1995}:
$d^6\underline{L}_{\eg}$).  A more recent study identified this high
spin state as $d^6$, but it is mixed with several other possible spin
states.\cite{Kunes2012a} Also the magnetic moment (ranging from
\SIrange{2.6}{3.19}{\muB})\cite{Lee2011, Wu2012} is reproduced by
theory and close to $\SI{3}{\muB}$, which agrees very well with the
intermediate spin state picture ($S=3/2$).  

However, there is only one paper, which discussed the Curie temperature
($\TC$) of \sco{3} from a theoretical point of view ($\TC\approx
\SI{1800}{\kelvin}$).\cite{Kunes2012a} This value is far too high in
comparison to the experimentally measured \add{values, which vary} from
\SIrange{212}{305}{\kelvin}.\cite{Takeda1972, Taguchi1978,
  Bezdicka1993, Kawasaki1996, Balamurugan2006prb, Long2011jpcm}
This variation might result from different synthesis techniques and
defects.  Usually, the cubic and stoichiometric phase \sco{3} is
prepared from the brownmillerite phase \sco{2.5}.\cite{Long2011jpcm}
The oxidation is either done by a high oxygen pressure during the
heating\cite{Long2011jpcm} or by electrochemical
oxidation.\cite{Bezdicka1993, Karvonen2007} Furthermore, most samples
are polycrystalline and the oxygen amount is smaller than the nominal
value ($\delta\geq0.05$).  The cubic structure is only stable for a
narrow range of $\delta$, \add{since \sco{3-\delta}} forms a homologous
series (\sco{(3n-1)/n}) for $\delta$ from 0.5 to 0 and changes its
structure from orthorhombic via cubic and tetragonal to cubic
again.\cite{Balamurugan2010jsnm,LeToquin2006, Karvonen2008cm}
Even for this cubic range close to $\delta=0$, the lattice constant
increases slightly with decreasing oxygen content, \cite{Karvonen2007,
  Balamurugan2006prb, Nemudry1996cm, LeToquin2006, Taguchi1979}
whereas the critical temperature decreases from
\SIrange{292}{182}{\kelvin}.  \cite{Balamurugan2006prb, Taguchi1979}
Anyway, these values for polycrystalline samples are smaller than for
a single crystal ($\TC=\SI{305}{\kelvin}$ and
$\delta=0.05$).\cite{Long2011jpcm} As for the $\TC$ values, the
measurements of the magnetic moments vary for different experiments
between \SIrange{1.2}{2.6}{\muB}.
\cite{Takeda1972,Bezdicka1993,Long2011jpcm}

\add{
Therefore, we studied in this work the electronic and 
magnetic properties of \sco{3-\delta} from \textit{ab initio}.
There were strong indications for the IS state 
proposed by Potze \etal,\cite{Potze1995} when using 
in the calculations more advanced exchange-correlation functionals other than 
the \add{generalized-gradient} approximation.
Thereby, the theoretically estimated critical temperature 
as well as its reduction with increasing amount of oxygen vacancies (\vo)
agreed well with the experimental observations.}

In the following, we start with a detailed description of \add{the studied structure} 
and the \add{applied} theoretical methods.  The results are discussed at 
first for the defect-free SCO. Afterwards, we extend our study
also to oxygen vacancies \add{in SCO.}  We close this work with a summary.

\begin{figure}
  \includegraphics[width=\columnwidth]{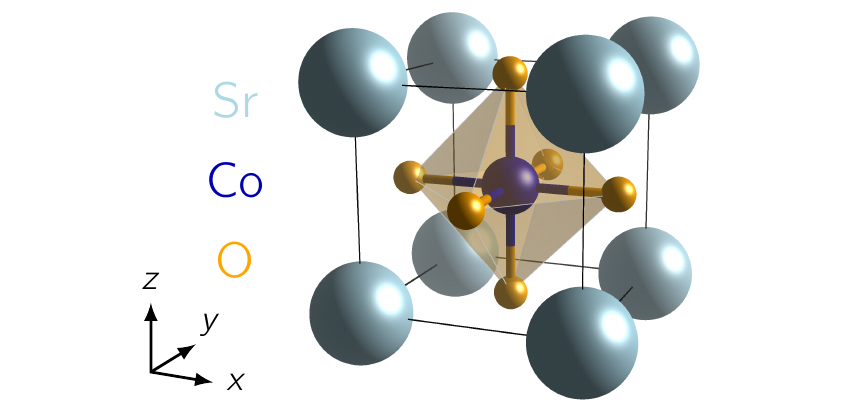}
  \caption{(Color online) Structure of the cubic phase of \sco{3}.}
  \label{fig:structure_sco}
\end{figure}

\section{Computational details}
\label{sec:comp details}
In this work, the calculations were performed for a primitive cubic
cell of \sco{3} (see Fig.~\ref{fig:structure_sco}).
\add{To study structural, electronic,
and magnetic properties of \sco{3}, we performed extensive
first-principles calculations within density functional theory
and combined two methods, which have been proven to be very
reliable in \add{this} particular context. The projector-augmented plane-wave method 
\add{of} the Vienna \textit{ab initio} simulation package \add{(VASP)}
\cite{Kresse1994prb, Kresse1996cms, Hafner2008jcc} was used for the 
calculation of total energies and \add{structural relaxation, to search 
for possible deviations from the cubic structure.} 
The main results were obtained with the Korringa-Kohn-Rostoker
Green's function (GF) method \add{HUTSEPOT}.
\cite{ [{HUTSEPOT is a Green's function method and developed by A. Ernst \etal at the Max Planck
Institute of Microstructure Physics, Germany. The basic features were described in }]
[{}]Luders2001jpcm}
In both methods, if not particularly stated otherwise, 
the \add{generalized-gradient} approximation
(GGA)\cite{Perdew1996prl} was used and we
considered the electronic correlation effects for the $d$ orbitals of
Co with an additional Hubbard $U$ within the GGA$+U$
approach\cite{Anisimov1991prb} in the
implementation of Dudarev \etal\cite{Dudarev1998prb} The $\Ueff=U-J$
was applied as a parameter varying from \SIrange{0}{9}{\electronvolt}.
}

\begin{figure}
  \includegraphics[width=\columnwidth]{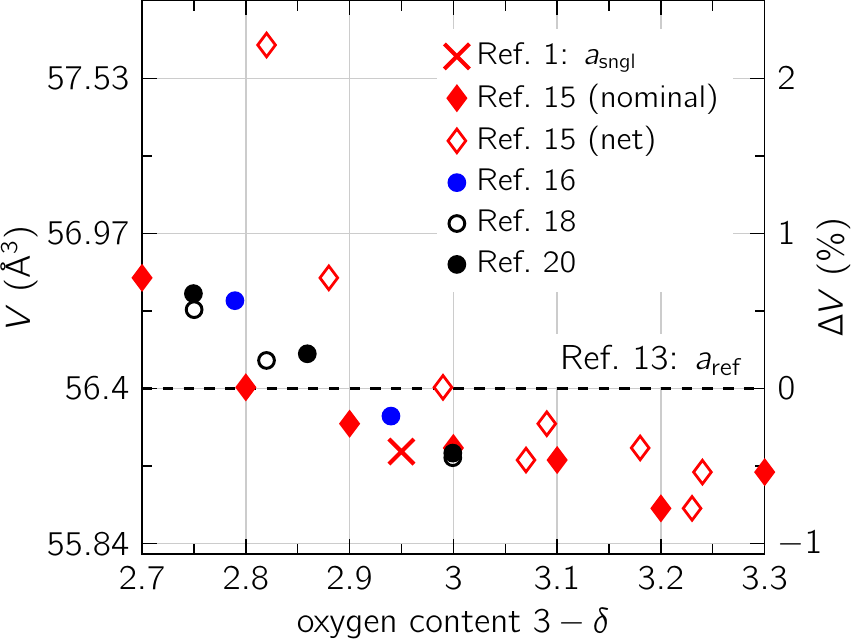}
  \caption{Experimental \add{lattice volume as a function of} oxygen
    concentration from several references. The change of the volume is
    \add{$\Delta V=(V-a_\text{ref}^3)/a_\text{ref}^3$ with respect to
    $a_\text{ref}=\SI{3.835}{\angstrom}$} (black dashed line). 
    The nominal and net
    values \add{of Ref.~\onlinecite{Balamurugan2006prb}}
    represent the provided and measured oxygen content.}
  \label{fig:V vs delta exp}
\end{figure}

\subsection{Structural relaxation}

The plane-wave
basis \add{for the VASP calculations} was taken with a cutoff energy of $\SI{460}{\electronvolt}$ and
a $\Gamma$-centered ($8\times8\times8$) $\vec{k}$-mesh was used.  It
reproduced the experimental unit cell volume of
Ref.~\onlinecite{Bezdicka1993} (cubic phase; $a_\text{ref}=\SI{3.835}{\angstrom}$) 
with \add{a deviation} of less than
\SI{3}{\percent} for the studied range of $\Ueff$. In the same order
of magnitude or smaller are \add{the differences (with respect to $a_\text{ref}$)} to the single-crystal
lattice constant\cite{Long2011jpcm}
$a_\text{sngl}=\SI{3.8289}{\angstrom}$ ($\Delta V=
\SI{-0.47}{\percent}$ with $V=(a_\text{sngl})^3$) or possible
variations due to oxygen vacancies observed e.g. in
Ref.~\onlinecite{Balamurugan2006prb} ($\Delta V=
+\SI{2.2}{\percent}$) or Ref.~\onlinecite{Karvonen2007} ($\Delta
V=+\SI{0.56}{\percent}$) (see overview in Fig.~\ref{fig:V vs delta
  exp}). \add{For the sake of completeness, we compared our \add{GGA results 
with hybrid functional calculations} using the common HSE03 functional, 
\cite{Heyd2003jcp} available within VASP but at the moment 
not available in the GF method. 
In the HSE03 functional, the mixing $\alpha= 0.25$ and the screening parameter 
$\mu = \SI{0.3}{\per\angstrom}$ are chosen for the Hartree-Fock exchange energy. 
Because of the high computational demands of such type 
of calculations the cutoff energy was reduced to \SI{400}{\electronvolt} and
the $\Gamma$-centered $\vec{k}$-mesh was only $4\times4\times4$. 
The lattice relaxation provided for a cubic phase the equilibrium lattice 
parameter $a_\text{HSE03}=\SI{3.833}{\angstrom}$, 
which is in a very good agreement with the experimental values shown above.
Although we did not observe any indication of a tetragonal distortion
in our GGA$+U$ calculations, we found by applying the HSE03 exchange functional
that the $a/c$ ratio deviated by few percent from one. 
Since the experiments for stoichiometric \sco{3} did not observed a tetragonal unit cell and
the electronic structure did not change substantially with respect to the cubic 
phase,} we stayed in the main part of the paper consistently with
\add{the cubic SCO and the lattice constant} $a_\text{ref}=\SI{3.835}{\angstrom}$
and investigated \add{a hydrostatic volume variation} separately in
Sec. \ref{sec:volume_effect}.

\subsection{Magnetic properties with the Green's function and Monte Carlo method}

\add{The calculations with the GF method were} performed
within the full charge density approximation, which takes into account
the non-sphericity of the charge density and improved the accuracy of
calculations for complex unit cell geometries.

\add{The GF method, which is based on the multiple scattering theory,
allows the calculation of the
magnetic exchange interactions $J_{ij}$ between the magnetic atoms at
site $i$ or $j$ by using the magnetic force theorem.\cite{Liechtenstein1987jmmm}}
\add{Therein, the magnetic moments at $i$ and $j$ of an ordered magnetic 
structure are rotated against each other about a small angle. From the resulting
energy variation follows}
\begin{align}
  J_{ij} = \frac{1}{8 \mypi} \int^{\EF}\!\!\!\!\diff\epsilon\, \text{Im}\, \trace_L
           \big(\Delta_i\scatpaopup{ij}\Delta_j\scatpaopdw{ji} +
                \Delta_i\scatpaopdw{ij}\Delta_j\scatpaopup{ji}\big)\mathcomma
  \label{eq:jij}
\end{align}
where $\scatpaop{ij}$ is the scattering path operator and
$\Delta_i=\singscatopup{i}-\singscatopdw{i}$ the difference between
the spin-dependent single scattering operators of site $i$. The trace
is taken over all relevant angular momentum quantum numbers $L=(l,m)$.

The $J_{ij}$ from \eqref{eq:jij} enter a classical Heisenberg Hamiltonian
\begin{equation}
  \hat{H}=-\sum_{i,j}J_{ij}\vec{S_i} \cdot \vec{S_j}\mathcomma
  \label{eq1}
\end{equation}
with the magnetic moments $\vec{S}_i$ and $\vec{S}_j$. The model
Hamiltonian \eqref{eq1} allows an estimate of the critical
temperature $\Tc$ with a Monte Carlo~(MC) simulation. Here, $\Tc^\text{MC}$
might mark the transition to any kind of magnetic ground state, which
might result from the calculations.  We \add{cross-checked} our results with
the mean-field approximation (MFA) and the well established random
phase approximation (RPA).~\cite{Tyablikov1995book} \add{All three methods
showed a similar tendency in the variation of $\Tc$ 
with respect to the considered influences like electron correlations or oxygen vacancies. 
Since the MFA is known to overestimate $\Tc$, 
we restricted the discussion to the MC results.}
For the particular MC simulations, we choose
\num{8000} magnetic atoms in a cluster with periodic boundary
conditions.  The temperature was reduced in steps of \SI{5}{\kelvin}
starting from a high-temperature disordered state above $\Tc^\text{MC}$.  At
every temperature $T$, we assumed that the thermal equilibrium was
reached after \num{20000} MC steps and after \num{20000} additional MC
steps thermal averages were calculated.  $\Tc^\text{MC}$ was then obtained from
the fitting of the temperature dependency of the magnetic
susceptibility, \add{cross-checked} by the temperature dependence of the
saturation magnetization and the heat capacity. The obtained
transition temperatures \add{were determined} within a numerical uncertainty range
of $\pm\SI{5}{\kelvin}$.  Further computational details of our MC
scheme can be found in
Refs.~[\onlinecite{Fischer2009prb,Otrokov2011prb,Otrokov2011prb-1}].
From the orientation of the magnetic moments at low temperatures and
the spin-spin correlation function, we deduced the magnetic ground
state of the simulated system, which was not only ferromagnetic (FM)
with the Curie temperature, $\TC$, as observed from the experiments,
but also antiferromagnetic (AFM) with the N\'eel temperature, $\TN$,
or a more complicated non-collinear ferri-magnetic spin arrangement
(FiM) with a transition temperature, $\Tc$, and potentially with a
saturation magnetization of zero.  We introduced those labels for the
different magnetic transitions for the sake of clarity throughout our
work and speak of $\Tc$ if the ground state is not clarified.

\add{In our calculations for a ferromagnetic ground state (reference state -- \rsfm), 
we found besides the magnetic cobalt ions small
induced moments at the oxygen sites.  Since those} moments disappear
usually at and above the magnetic transition \add{temperature, their}
magnetic coupling can lead to a wrong estimation of the critical
temperature.  \add{We compared the results of \eqref{eq:jij} for the FM reference
state \rsfm} with the magnetic coupling
constants obtained at a high temperature paramagnetic (PM) reference state (\rspm). 
Such state can be
modeled successfully with the disordered local moment (DLM)
theory\add{ using the GF method.}\cite{Staunton1984jmmm, Gyorffy1985jpf} 
Here, arrangements of
local magnetic moments $\{\vec{S}_i\}$ at the sites $i$ are thought to
fluctuate independently.  Above $\Tc$, the orientations of those local
moments are randomly distributed and the average magnetization per
site is zero. Hence, the induced moments vanish.  From the
computational point of view, the coherent potential approximation
(CPA) \cite{Soven1967pr} as it is implemented within the multiple
scattering theory\cite{Gyorffy1972prb} can be used to model the
susceptibility or the electronic structure for such disordered
magnetic systems.  \cite{Staunton1984jmmm, Gyorffy1985jpf} With
respect to the calculation of the $J_{ij}$ in \eqref{eq:jij}, the
scattering path operators $\scatpaop{ij}$ of the perfect magnetic ordering
will be exchanged with the ones of the effective CPA
medium $\scatpaop{ij}_\text{C}$.  The random orientation of the
magnetic moments in the DLM picture causes usually an increase in the
size of the orbitals.  So, the magnetic coupling is usually stronger
in case of a DLM calculation since the strength of the magnetic
coupling \add{is proportional to the overlap between the contributing
orbitals.} This means for the resulting transition temperature that it
can become either larger or smaller depending on the type of magnetic
exchange, e.g. a reduction for antiferromagnetic (super-) exchange.

To describe the \add{oxygen-deficient \sco{3-\delta} in the GF method}, 
we applied also the CPA
using a certain amount ($\delta$) of empty spheres at the oxygen
sites to mimic \vo.  However, \add{vacancies} may lead to
substantial relaxations of the underlying crystal structure. We did
not account for such structural deformations in our CPA calculations
but investigated their impact on the magnetic interaction using 
a supercell approach \add{with the GF method.} 
We found only minor changes in the exchange
constant values and, therefore, the discussion below reports results
from the ideal cubic structure.

\section{Results}
\subsection{Defect-free SCO}

\begin{figure}
  \includegraphics[width=\linewidth]{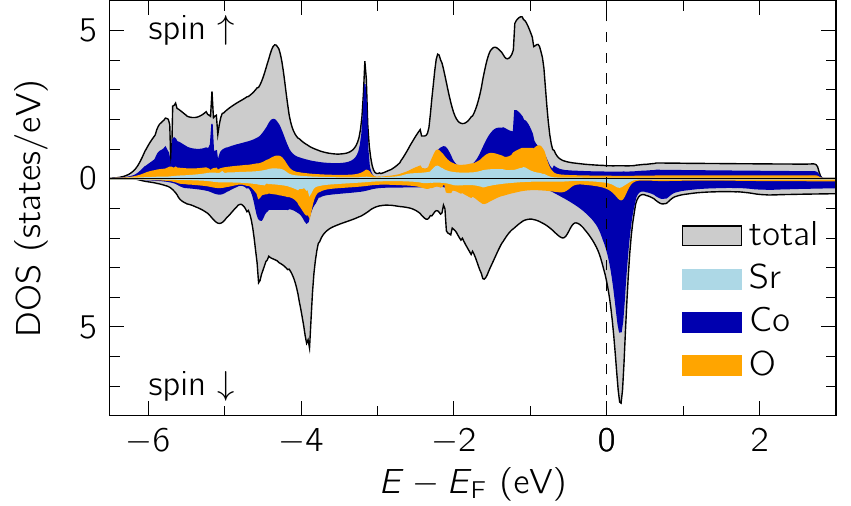}
  \caption{(Color online) Atomic and spin-resolved density of states
    of \sco{3} with contributions of each species and spin-up (upper
    panel) and spin-down (lower panel) obtained \add{with the GF method (GGA).} The
    oxygen $p$ states at all three oxygen ions in the unit cell are
    three fold degenerate (only one is shown). 
    \add{Main feature of the LDOS are similar in VASP (not shown).}}
  \label{fig:LDOS_sco_U=0}
\end{figure}

We investigated at first the electronic
properties of \sco{3} for the experimental lattice constant.  Due to
the oxygen octahedron, which surrounds the Co ion, the $d$ states of cobalt
experience a crystal field splitting of cubic symmetry, which results
in three $\tg$ and two $\eg$ degenerated states. The \add{orbitals
corresponding to the} $\eg$ states are
oriented along the coordinate axes pointing to the oxygen ions and 
\add{those corresponding to} the
$\tg$ states are pointing to the next nearest \add{neighboring Co ions} (see
Fig.~\ref{fig:structure_sco}).

The local density of states (LDOS) within the GGA shows 
\add{for VASP and HUTSEPOT} an almost
fully occupied majority spin channel and a pronounced peak of the Co
$\tg$ spin-down states at the Fermi energy $\EF$ (see 
Fig.~\ref{fig:LDOS_sco_U=0}). All other cobalt $3d$
states, $\due$, $\dut$ and $\dde$, are below the $\EF$ and
smeared over a large energy range due to a strong hybridization with
the $p$ states of oxygen. 
\add{This is in a good agreement with
previous results\cite{Kunes2012a} but contradicts the
IS state picture with
particular occupied and unoccupied $\tg$ Co states ($\tg^4\eg^1$).
\cite{Potze1995}}

\add{The total moment of this IS state model would be
theoretically \simB{3} ($s=3/2$). 
Although the total magnetic moment of $\mu=\simB{2.281}$ 
calculated with the GF method and the GGA functional
was smaller than in the IS state model, 
it agreed well with the range of the experimentally observed magnetic moments
($\mu\approx\simB{2}$ to \simB{2.5}).
The main contribution to the total moment 
originated correctly from the Co ions with only small 
induced moments of \simB{0.15} and
\simB{0.04} at the oxygen and strontium ions, respectively.
The discrepancy between the IS state model and the experiments
was always attributed to possible defects.
However, when the IS state model is valid,
theoretical calculations of a defect-free SCO 
should reproduce the total magnetic moment.} 

\begin{figure}
  \includegraphics[width=\linewidth]{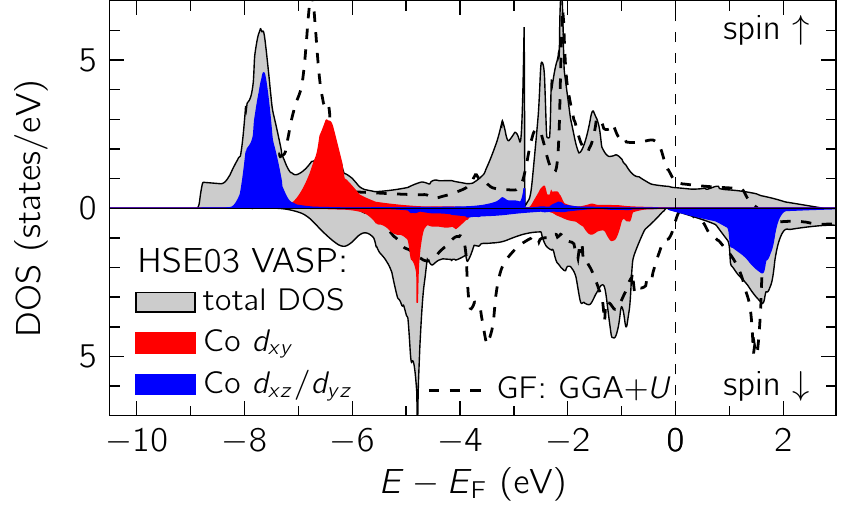}
  \caption{(Color online) \add{Spin-resolved density of states
    of \sco{3} \add{in the spin-up (upper panel) and spin-down channels} (lower panel) 
    obtained with VASP (HSE03 functional) and
    GF method (GGA$+U$ with $U=\SI{5}{\electronvolt}$). Only
    the non degenerated Co $3d$ states are shown.}}
  \label{fig:LDOS_sco_VaspHSE_KkrGGA+U5}
\end{figure}

\add{Possible shortcomings might originate from the GGA electron correlation functional,
which often lacks a sufficient description of localized transition metal $d$ states. For comparison,
we calculated the total magnetic moment and the DOS of SCO also with the hybrid functional
HSE03 available in VASP. The obtained total magnetic 
moment of \simB{2.9} within VASP and the HSE03 functional agreed well with the IS state model.
Additionally, the DOS showed also the expected orbital occupation $\tg^4\eg^1$
of the Co states: two unoccupied $\tg^\downarrow$ ($d_{xz}$ and $d_{yz}$ above $\EF$)
and one occupied $\tg^\downarrow$ state ($d_{xy}$ below $\EF$) 
(see Fig.~\ref{fig:LDOS_sco_VaspHSE_KkrGGA+U5}).}

We have to note that due to the symmetric cubic structure, 
the particular localization of $\ddxy$ is
arbitrary and depends on the starting point of the self-consistent
calculation.   Another localized
Co $d^\downarrow$ state is also possible and was observed during the
calculations. However, for a consistent description we continued
throughout this work with one particular configuration (the singlet
$\ddxy$ state).

\begin{figure}
  \includegraphics[width=\columnwidth]{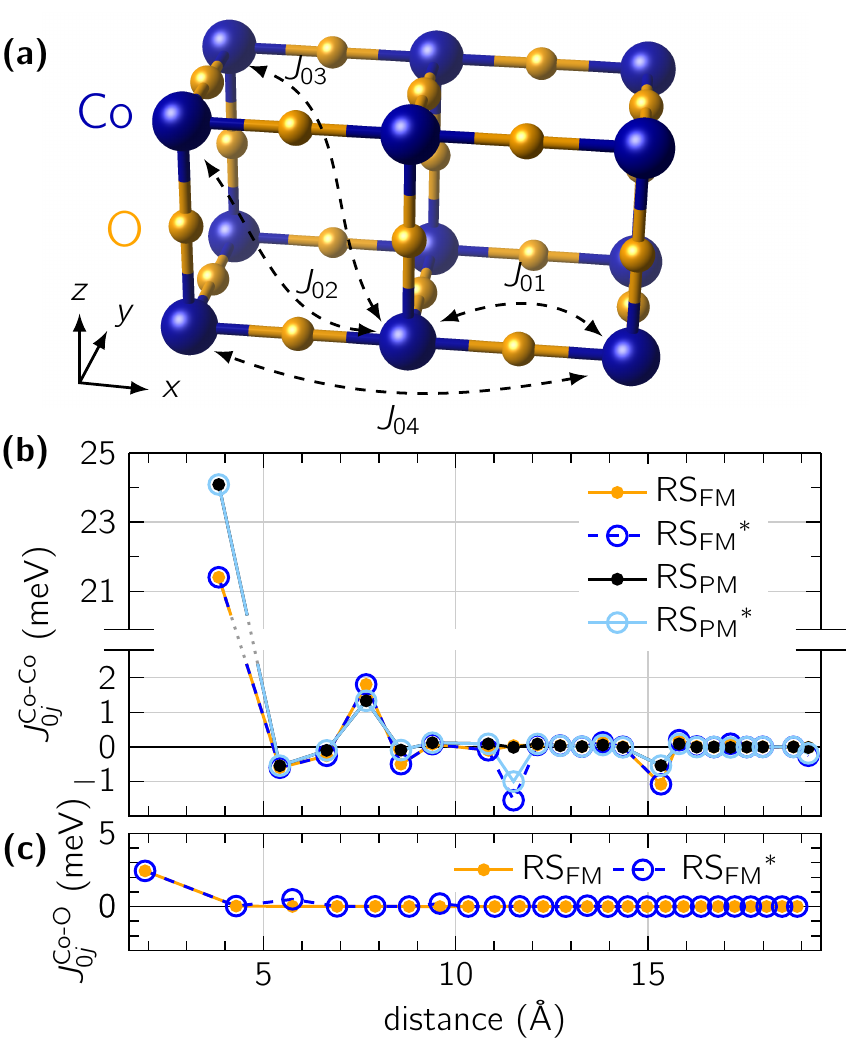}
  \caption{(Color online) (a) Schematic view of the orientation of the
    first magnetic exchange interactions $J_{ij}$ only between the Co ions.
    The non magnetic Sr ions are not shown.  The notation $J_{01}$,
    $J_{02}$, $\ldots$ means nearest neighbors (NN), next NN, etc.
    Below, the values for the coupling between Co-Co and Co-O are
    given in (b) and (c), respectively. Both calculated with 
    \add{the GF method (GGA) for a FM and
    PM (DLM model) reference state, respectively.}  The asterisk marks those
    coupling constants \add{which strictly follow only one direction in space
    (along a single cube edge in (a)).}  The abscissa in
    (b) is non-continuous, due to the large differences in the
    values.}
  \label{fig:U=0_Jij_vs_distance}
\end{figure}

\add{In order to calculate the critical temperature, 
the magnetic exchange interactions for the nearest
neighbor atoms were calculated with the GF method and the GGA functional.
For \rsfm the magnetic interaction parameters show mainly a ferromagnetic (positive)
coupling:} strong between two adjacent Co ions and much weaker
between the Co ions and the induced magnetic moments of the
surrounding oxygen ions (see Fig.~\ref{fig:U=0_Jij_vs_distance}).  The
coupling to the Sr ions was one order of magnitude smaller and was
ignored in the following discussions.  After the 8th shell
($d=\SI{7.7}{\angstrom}$), most of the coupling constants decay fast
while other long-range interactions reflect the metallic character of
\sco{3}. Along the straight exchange paths (along a particular
coordination axis, also marked with an asterisk in
Fig.~\ref{fig:U=0_Jij_vs_distance}) the coupling via the oxygen ions
remains stronger.  The same tendency can be observed for the
coupling between the Co ions in the DLM model \add{($J_{ij}$ calculated with \rspm)}.  Only the nearest
neighbor interactions became larger while the other coupling constants
were reduced.  \add{At the oxygen ions, the induced magnetic moments
vanished in the DLM model and no exchange interactions were found between them.}
In the Monte Carlo method, we took into account all coupling constants up to the distance
of \SI{15.34}{\angstrom}. \add{The resulting Curie temperature 
did not agree with the experimental results (within the DLM model
$\TC^\mathrm{MC}=\SI{771}{\kelvin}$ compared to $\TC^\text{exp}\approx\SI{280}{\kelvin}$).}
\add{The reason for this high $\TC^\text{MC}$
might be an overestimation of the magnetic coupling between the Co atoms.}
The \add{increase in the localization} of $3d$ states 
\add{by using electron correlation corrections or more advanced functionals}
should decrease as well the orbital overlap and thereby reduce the exchange
coupling. 

\add{In summary, the DOS, the total magnetic moment and the critical 
temperature, illustrate the need of electron correlation corrections
for the description of the properties of SCO. 
Unfortunately, the calculations with 
hybrid functionals consume large amounts of 
computational resources and were also 
at the moment not available within HUTSEPOT,
which was needed for the calculations of the $J_{ij}$.}
However, the parametric GGA$+U$ approach was less time consuming
and \add{provided similar} results, as it is discussed 
\add{in the following Sec.~\ref{sec:correlation correction}.
For that reason, we stayed in the remaining of this work with the GGA$+U$ method.
Additionally, we describe also the oxygen-deficient material,
where we expect a reduction of the critical temperature and a slightly
enhanced volume.\cite{Balamurugan2006prb}}
%
%
\subsection{\add{Electronic correlation corrections}}
\label{sec:correlation correction}
\begin{figure}
  \includegraphics[width=\columnwidth]{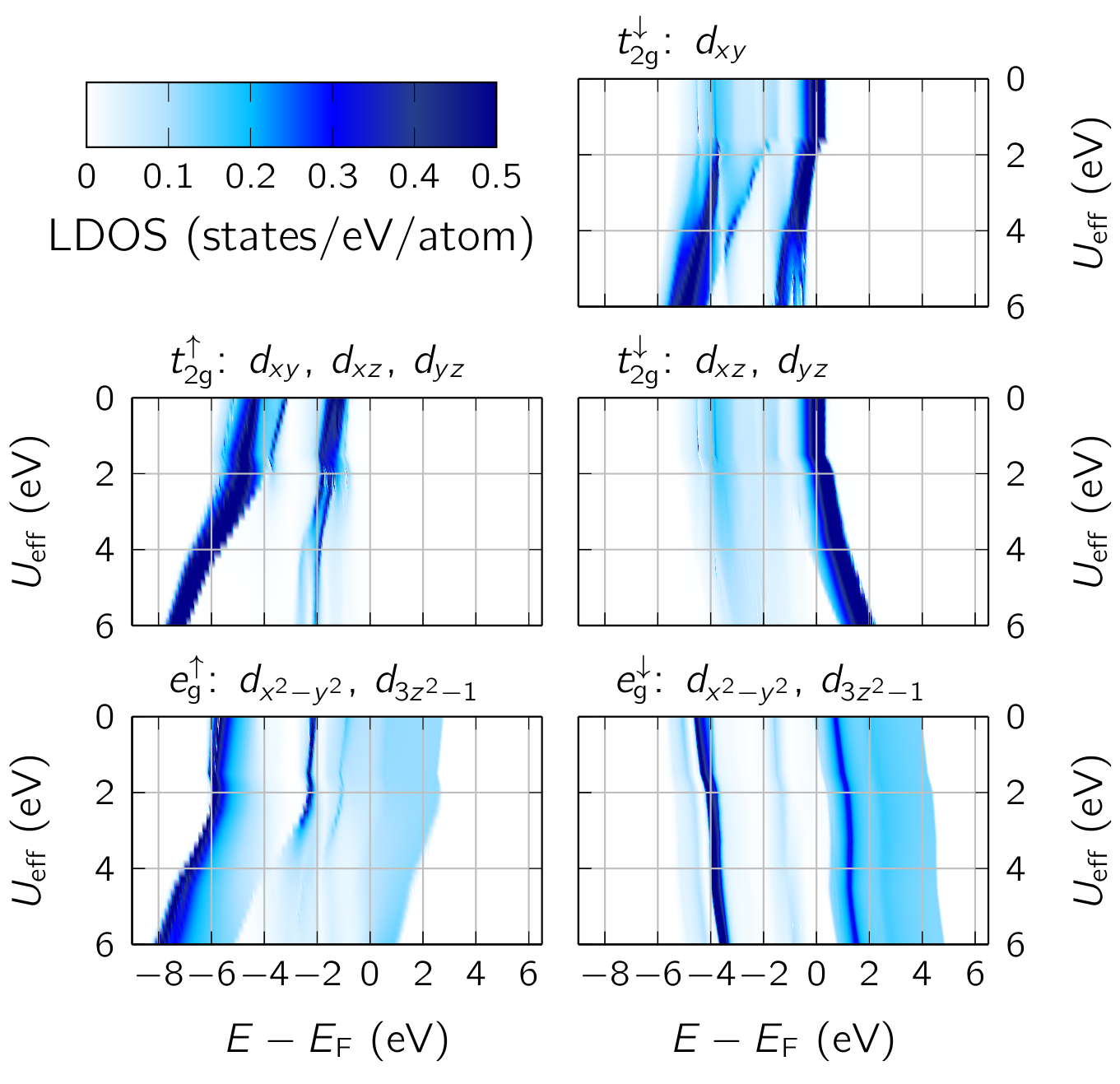}
  \caption{(Color online) Contour plots of the local density of states
    (LDOS) in the ground state for the Co $d$ states in dependence of
    the correlation parameter $\Ueff$ (ordinate) \add{obtained with the GF method.} 
    They are spin-resolved (\add{left:} up, \add{right:} down) and collected according to their
    initial degeneracy. For higher $\Ueff$ (not shown), the energy
    shift continues and there are no significant changes to observe.}
 \label{fig:LDOS_sco_vs_U}
\end{figure}

The common way to consider the electronic correlations \add{within the GGA$+U$ method}
is to optimize
the value of the repulsive $\Ueff$ with respect to the experimental
data \add{for} structural and magnetic properties of the system. As we see
from former studies, the value of $\Ueff$ might range from
$\SI{2.5}{\electronvolt}$ (Ref.~\onlinecite{Lee2011}) to
$\SI{8}{\electronvolt}$ (Ref.~\onlinecite{Wu2012}). The constrained
random-phase approximation provided a value of
$U=\SI{10.83}{\electronvolt}$ and $J=\SI{0.76}{\electronvolt}$ for the
Co $d$ states\cite{Kunes2012a} which seems to be \add{too} high for a
metallic system.

Since the correct value of $\Ueff$ is hardly to estimate from
first-principles and the above reference values scatter quite a bit,
we investigated the electronic structure and the occupation of the Co
$d$ states for the whole range of $\Ueff$ from \sieV{0} to
\sieV{9} \add{with the GF method} (see Fig.~\ref{fig:LDOS_sco_vs_U}).
Interestingly, for the first few steps of the calculations
($\Ueff\leq\SI{1.5}{\electronvolt}$) the $d$ states preserve their
degeneracy in $\tg$ and $\eg$ \add{states}.  At the Fermi energy $\EF$, the large
peak of the Co $\tg$ states does not move due to an interplay of the
Coulomb exchange and the crystal-field energy.  Only for a larger
$\Ueff$, the degeneracy is lifted
\add{as it was observed before in the DOS 
calculated with the HSE03  (see Fig.~\ref{fig:LDOS_sco_VaspHSE_KkrGGA+U5}). A singlet state ($\ddxy$)} becomes
occupied while the doublet ($\ddxz$ and $\ddyz$) is pushed above $\EF$.  On the
other hand, in the \add{spin-up} channel the orbitals remain \add{degenerate} for
the whole range of $\Ueff$ and become strongly localized (see much
higher contrast for $\duxy$ in Fig.~\ref{fig:LDOS_sco_vs_U}).  
\add{It matches well with $\tg^4\eg^1$ and the IS state model.}

\add{A similar loss of degeneracy in the $3d$ states was also observed in calculations with VASP 
and GGA$+U$, but at higher values for $\Ueff$ than in the GF method (happening between \SIrange{5}{7}{\electronvolt}, 
not shown).}
\add{Furthermore, we obtained a very good agreement of the 
electronic structure around the $\EF$, calculated either with the GF method at $\Ueff=\SI{5}{\electronvolt}$ 
or with VASP (HSE03) (see Fig.~\ref{fig:LDOS_sco_VaspHSE_KkrGGA+U5}).
For states further below $\EF$, 
the differences become larger but those states contribute
only little to the orbital overlap needed for the calculation
of magnetic exchange parameters with the magnetic force theorem.}

\begin{figure}
  \includegraphics[width=\columnwidth]{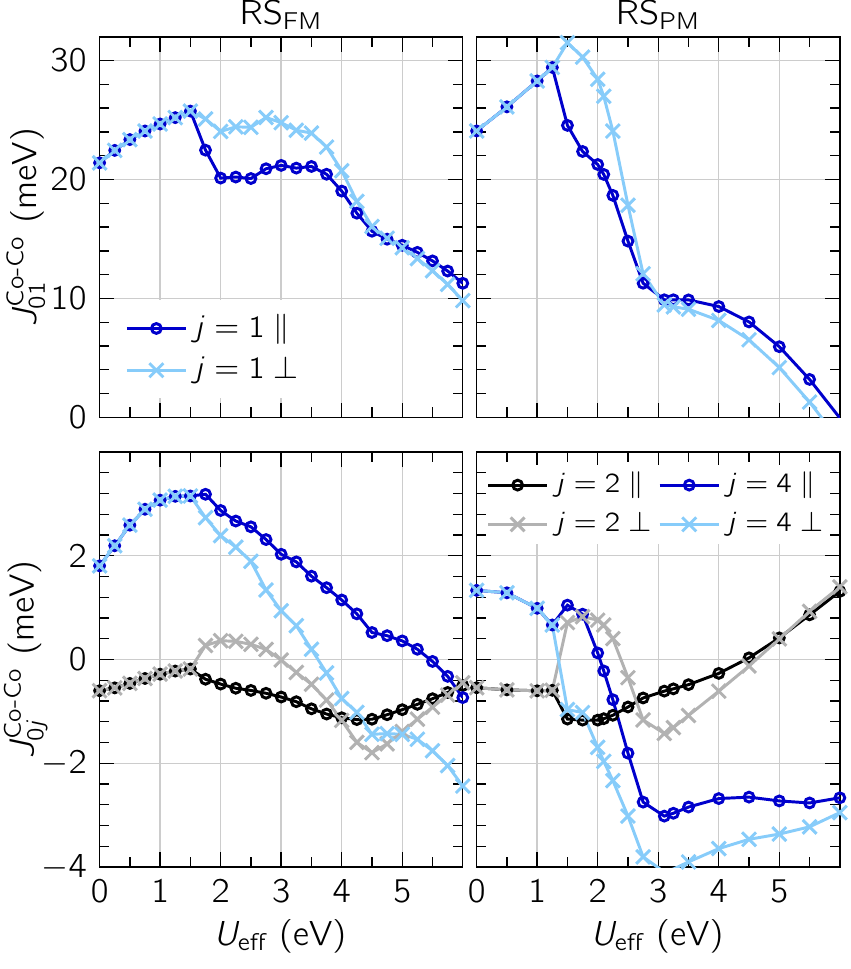}
  \caption{(Color online) The magnetic exchange interactions
    $J_{0j}^\text{Co-Co}$ \add{as a function of} $\Ueff$ 
    \add{for a ferromagnetic (left) or 
    paramagnetic (right) reference system.}  
    They \add{are subdivided} into two groups, $\parallel$ and $\perp$, for
    $J_{0j}^\text{Co-Co}$ only in the $x$-$y$-plane and those with
    contributions also in $x$ direction (directions given in
    Fig.~\ref{fig:U=0_Jij_vs_distance}(a)).  The
    $J_{03}^\text{Co-Co}$ (not shown) does not split due to $\Ueff$
    and are always small ($<\simeV{0.3}$). }
  \label{fig:magnetic_exchange_interactions_vs_U}
\end{figure}

For the understanding of the critical temperature, we calculated 
\add{with the GF method} the
magnetic coupling parameters $J_{ij}$ \add{for the 
FM and PM (DLM model) reference state, respectively
(see Fig.~\ref{fig:magnetic_exchange_interactions_vs_U})}. 
Both reflect the change in
the degeneracy of the Co $d$ state in the LDOS.  Due to the splitting
of the $\tg^{\downarrow}$ states into \add{the degenerated doublet 
($d_{xz}^{\downarrow}$ and $d_{yz}^{\downarrow}$)} and a \add{singlet
($d_{xy}^{\downarrow}$),} the $J_{ij}$ with either ($\parallel$)
in $x$ and $y$ direction or ($\perp$) in $z$ direction were different
for $\Ueff>\SI{1.5}{\electronvolt}$.  
For the ground state calculation (\rsfm), the most dominant
coupling constants, $J_{01}^\text{Co-Co}$ and $J_{04}^\text{Co-Co}$,
are strongly ferromagnetic (see left hand side in
Fig.~\ref{fig:magnetic_exchange_interactions_vs_U}).  Both
interactions are mediated by oxygen ions between Co ions forming
either a Co-O-Co or a Co-O-Co-O-Co chain (see inset in
Fig.~\ref{fig:U=0_Jij_vs_distance}).  Those bonds connect mostly the O
$p$ states with the Co $\eg$ states ($\sigma$ bonds).  This typically
antiferromagnetic (AFM) superexchange is suppressed by the
metallic character of \sco{3} and we observe band magnetism.  In the
degenerated parameter region, the coupling becomes even stronger with
$\Ueff$ until \SI{1.5}{\electronvolt} because the Co $\due$ and $\dde$
states are either pushed below or above $\EF$,
respectively. This increases the exchange splitting and, therefore,
the magnetic coupling.  In contrast, the coupling between two Co ions
enclosing a \SI{90}{\degree} angle ($J_{02}^\text{Co-Co}$) is small
and AFM, while the next coupling ($J_{03}^\text{Co-Co}$) is very weak
compared to the other interactions.

After that, in the symmetry broken regime, the competing superexchange
overcomes the band magnetism and we observed an increasing
localization of the $d$ states (see stronger contrast in
Fig.~\ref{fig:LDOS_sco_vs_U}).  This reduces, in general, the overlap
of the orbitals and the magnetic exchange interactions. It is visible
e.g. for $J_{01}$ and $J_{04}$ in
Fig.~\ref{fig:magnetic_exchange_interactions_vs_U} but also for other
$J_{0j}$ (not shown).  On the other hand, the modifications in the
coupling constants are much more complex due to the changing LDOS.
However, some simple tendencies can be observed, e.g. due to the
localization of the Co $\dxy$ states, the magnetic coupling for
$\Ueff<\SI{4}{\electronvolt}$ in the $x$-$y$-plane ($\parallel$
contributions in $J_{01}$) becomes smaller than the out-of-plane
($\perp$) contributions. The $\perp$ parts of $J_{02}^\text{Co-Co}$
even change their character from AFM to FM.
Another significant change for all coupling constants is visible at
$\Ueff=\SI{4.5}{\electronvolt}$ and might be correlated with vanished
states in the $\dxy$ LDOS indicated by a loss of contrast (see
Fig.~\ref{fig:LDOS_sco_vs_U}). At the end of the shown range, the
strength of the nearest neighbor magnetic exchange coupling was only
half of its starting value. It reduces further, for even higher
$\Ueff$ (not shown) and leads to an undesired antiferromagnetic ground
state.

The \add{overall} tendencies for the \add{\rsfm calculations} were in
general also observed in the DLM picture (\rspm), although the changes were much
stronger, e.g.  at $\Ueff=\SI{6}{\electronvolt}$ the $J_{01}$ were
reduced to zero and $J_{04}$ is strongly AFM. Furthermore, the loss of
degeneracy was visible already for smaller $\Ueff$ (see
Fig.~\ref{fig:magnetic_exchange_interactions_vs_U}).  Both changes are
explained by the larger extent of the Co $d$ orbital due to the random
distribution of the magnetic moments in the DLM theory as already
stated in section~\ref{sec:comp details}.  The increasing orbital
overlap enhances, on the one hand the AFM superexchange and on the
other hand, alters the competition between the crystal field and
Coulomb energy which restored the degeneracy for small $\Ueff$.

\begin{figure}
  \includegraphics[width=\columnwidth]{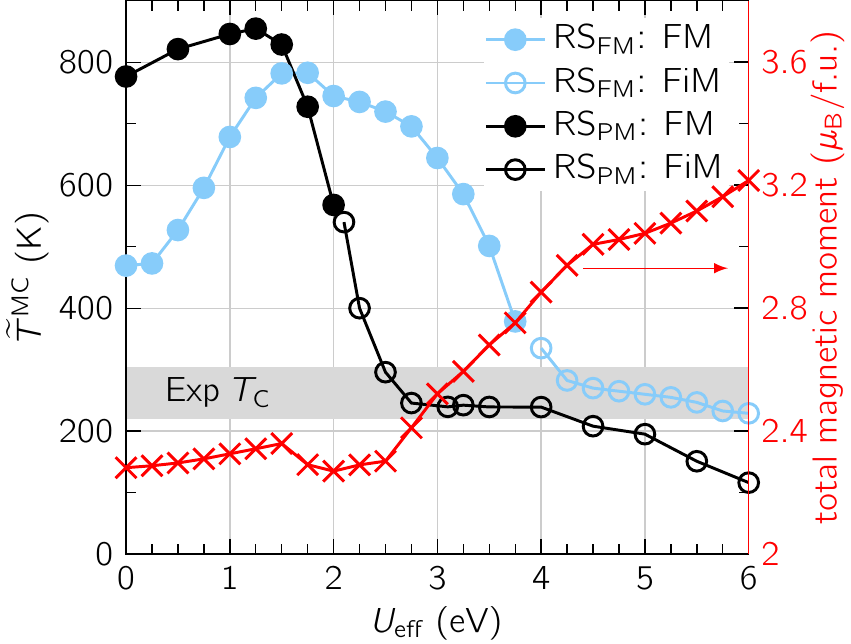}
  \caption{(Color online) The critical temperatures for the magnetic
    transition (left axis) and the total magnetic moment per
    functional unit \add{at the ground state (right axis) as a function of} 
    $\Ueff$. The Monte Carlo calculations show a change in the
    magnetic ground state from clear ferromagnetic behavior to a FiM
    situation at $\Ueff\approx\SI{2.1}{\electronvolt}$ and
    \SI{4}{\electronvolt} for the \add{\rspm and \rsfm}
    calculations, respectively (see text). The gray shaded area
    indicates the experimental $\TC$ range. }
  \label{fig:critical_temperature_vs_U}
\end{figure}

The critical temperatures \add{obtained from the Monte Carlo simulation
using the magnetic exchange parameters in the \rsfm or with the 
DLM model follow in general a similar tendency as the nearest neighbor 
coupling constant} $J_{01}^\text{Co-Co}$ (see Fig.~\ref{fig:critical_temperature_vs_U}).
\add{They show for both sets of magnetic coupling parameters} a linear increase up to
$\Ueff=\SI{1.5}{\electronvolt}$.  The critical temperatures obtained
with the $J_{0j}(\text{\rsfm})$ remained around $\SI{750}{\kelvin}$
with increasing $\Ueff$ up to $\approx\SI{3}{\electronvolt}$, and
drop down sharply in the following while the DLM results decrease
linearly immediately above \SI{1.5}{\electronvolt}. \add{So,  
$\Tc^\text{MC}(\Ueff)$ calculated for \rspm reaches the experimentally relevant range 
already for a smaller $\Ueff=\SI{2.5}{\electronvolt}$ than for \rsfm (\SI{4.5}{\electronvolt}). 
A reason for the smaller $\Ueff$ was already discussed above for the $J_{ij}$ --
the larger overlap of the orbitals in the more realistic DLM model. 
Additionally, the  \rspm calculation showed after the kink an almost constant 
$\Tc^\text{MC}(\Ueff)=\sikelvin{240}$ for a larger range of $\Ueff$ parameters (\sieV{2.75} to \sieV{4}). 
On the other side,
the ground state calculation at \rsfm with the GF method (GGA$+U$) returns
for $\Ueff\approx\SI{5}{\electronvolt}$ 
a $\Tc^\text{MC}$ inside the experimental range.
This observation matches well with the above discussion about the
electronic structure and the correspondence of GF method (GGA$+U$) and VASP (HSE03)
calculations at $\Ueff\approx\SI{5}{\electronvolt}$ (see Fig.~\ref{fig:LDOS_sco_VaspHSE_KkrGGA+U5}).}

Furthermore, we account in Fig.~\ref{fig:critical_temperature_vs_U}
for the magnetic ground state observed in the Monte Carlo study with
different symbols, having either a FM or FiM ground state.  The DLM
theory predicted only a FiM \add{ordered} ground state 
for the range of
$\Tc^\text{MC}$ ($\Ueff\approx\SI{2.5}{\electronvolt}$) equivalent to
experimental results. This is, however, still physically reasonable.
While at the critical temperature the induced moments at oxygen might
be zero, they will appear at lower temperatures. This leads to a
ferromagnetic ground state obtained with the $J_{ij}(\text{\rsfm})$
for even larger $\Ueff$ (see blue curve in
Fig.~\ref{fig:critical_temperature_vs_U}).

On the other hand, the total magnetic moment increases monotonously in
the whole $\Ueff$ range with different linear slopes (see red line in
Fig.~\ref{fig:critical_temperature_vs_U}). Only for changing
degeneracy for $\Ueff>\SI{1.5}{\electronvolt}$ the total moment
decreases slightly due to a reduction of the induced moments to
\SI{0.11}{\muB} for the two oxygen ions (O$_\text{x}$ and
O$_\text{y}$), which lie in the same $x$-$y$-plane as the Co ions.  In
the following, the linear slope changed around
$\Ueff\approx\SI{2.5}{\electronvolt}$ and
$\Ueff\approx\SI{4.5}{\electronvolt}$,  \add{which correspond to
disappearing peaks in the LDOS (see Fig.~\ref{fig:LDOS_sco_vs_U}). 
At the latter $\Ueff$, the total moment is 
\simB{3} and matches well on the one hand side with the IS state model
and on the other hand with the HSE03 calculation in VASP.}

\add{We found in summary that the PM model (DLM theory) 
yields for the critical temperature a good qualitative agreement 
with the measurements\cite{Long2011jpcm,Balamurugan2006prb} only for a small
correlation parameter of $\Ueff=\SI{2.50}{\electronvolt}$ to
\SI{2.75}{\electronvolt}.}
Still, a exact comparison \add{with the measured $\TC$ remained} complicated
due to the different experimental setups, single-crystalline or
polycrystalline samples, varying growing techniques or different
oxygen content.

%
%
\subsection{Effects of oxygen vacancies}
We used \add{the CPA of the GF method} to substitute the
oxygen sites with a certain concentration of empty spheres, modeling
the oxygen vacancies.  For \add{a low oxygen deficiency in \sco{3-\delta} up to
\SI{5}{\atper},} there is no experimental evidence for an ordering of \vo. 
So, we assumed randomly distributed oxygen vacancies which
matches well with the concept of the CPA.

\begin{figure}
  \includegraphics[width=\columnwidth]{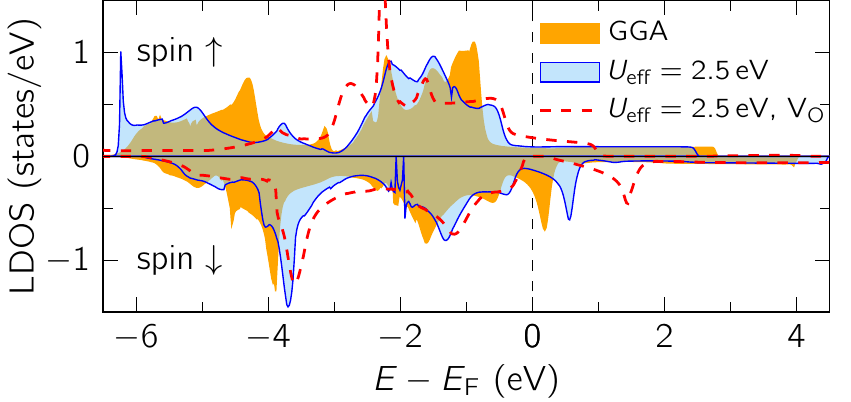}
  \caption{(Color online) LDOS of the oxygen along the $z$ direction
    (O$_{z}$) for GGA, $\Ueff=\SI{2.5}{\electronvolt}$ and with
    \SI{5}{\atper} oxygen vacancies (\vo).}
  \label{fig:oxygen_LDOS}
\end{figure}

The electronic structure of the oxygen ions is mainly dominated by the
strong hybridization with the Co $d$ states
(see Fig.~\ref{fig:LDOS_sco_U=0}).  By including of few atomic percent of oxygen vacancies
(\SI{5}{\atper}) in SCO, the unoccupied peak above $\EF$ is
shifted to higher energies (see Fig.~\ref{fig:oxygen_LDOS}),  \add{a similar 
effect as increasing $\Ueff$ (see GGA results for comparison in Fig.~\ref{fig:oxygen_LDOS}).
Both lead to an enhancing orbital localization -- either removing effectively oxygen from 
the lattice or reducing the electron hopping between the Co ions. 
Anyway, these unoccupied oxygen states} can
be interpreted as the ligand hole of $\eg$ symmetry, which is expected
for the IS state $\tg^4\eg^1$.\cite{Potze1995} It is stabilized by the ligand hole state
$d^6\underline{L}_{\eg}$ where the hole couples antiferromagnetically
to another $\eg^\uparrow$ in $d^6$: $\tg^4\eg^2$.  We found this
configuration also in our LDOS calculation with almost fully occupied
orbitals for $d^\uparrow$ and $\tg^\downarrow$ (see \ref{fig:LDOS_sco_vs_U}). 
\add{Although our} method is not directly
comparable with the dynamic mean-field theory (DMFT) method of
Kune\v{s} \etal\cite{Kunes2012a}, we note that the same spin
configuration appeared also in their calculation with the highest
multiplet weight.

\begin{figure}
  \includegraphics[width=\columnwidth]{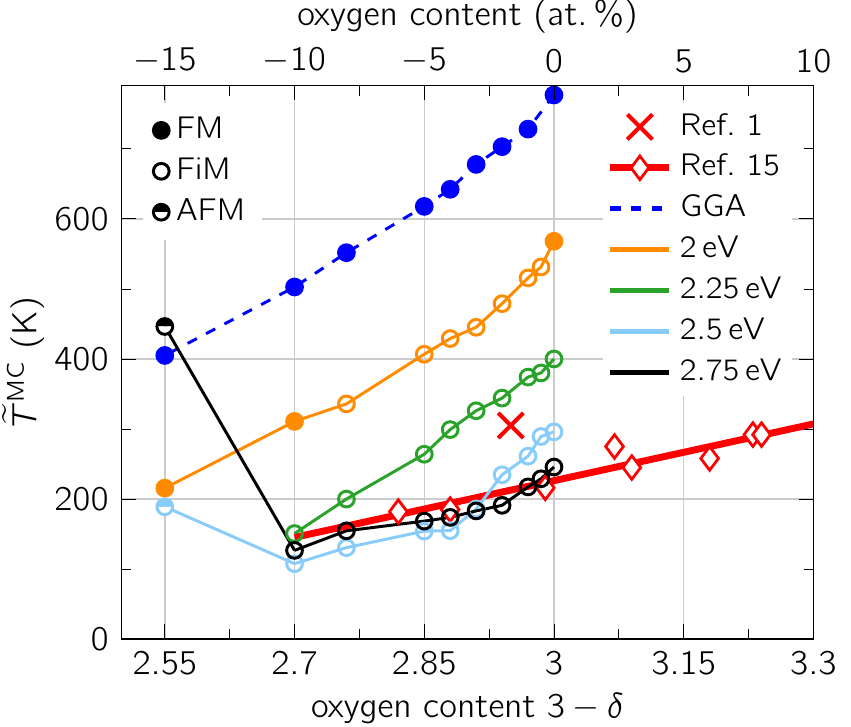}
  \caption{(Color online) \add{The theoretical critical temperature (DLM model) in
    dependence of the oxygen content in \sco{3-\delta} compared with
    experimental references. The GGA calculations shown as dashed blue line 
    and GGA$+U$ calculations shown as differently colored solid lines.} 
    The open and full circles \add{in the corresponding colors
    indicate the calculated data points and} the magnetic ground state as shown in
    Fig.~\ref{fig:critical_temperature_vs_U}.  An additional
    antiferromagnetic state (AFM) is marked with a half filled circle
    (see text).}
  \label{fig:critical_temperarture_vs_vacancies}
\end{figure}

\add{We calculated again the magnetic transition temperature $\Tc^\text{MC}(\delta)$
based on the \textit{ab initio} magnetic exchange coupling constants in \rspm
with the MC method but included different amounts of 
oxygen vacancies $\delta$. Although we observed at the relevant critical temperatures no
ferromagnetic ordering, calculations with the $J_{ij}$ in \rsfm revealed a
ferromagnetic ground state at low temperatures for all considered
vacancy amounts \add{(not shown)}. The results 
were compared with the experimental measurements.\cite{Long2011jpcm,Balamurugan2006prb}
The experiments with polycrystalline samples\cite{Balamurugan2006prb}
varied the oxygen content by $\pm\SI{6}{\atper}$ with respect
to the stoichiometric sample and found decreasing $\TC$ with 
reduced oxygen amount (see
Fig.~\ref{fig:critical_temperarture_vs_vacancies}).  A positive value
means excess oxygen, which might occupy unknown interstitial sites or
form more complicated point defects.  This is beyond our aims and we
restricted the current study to model only \add{oxygen deficiency} by the
introduction of oxygen vacancies into the cubic unit cell. In contrast,
the single-crystalline $\TC$ was found to be remarkably higher, but 
the lack of more data points complicates the comparison.\cite{Long2011jpcm}}

\add{In theory, we expected also a reduction of $\Tc^\text{MC}(\delta)$,
since we found that the localization of the orbitals through oxygen vacancies
was like applying an $\Ueff$ parameter. Therefore, the magnetic interactions
in the Co-O-Co bonds were weakened as well. This expectation 
was generally fulfilled for the calculated $\Tc^\text{MC}(\delta)$ at 
different $\Ueff$: the GGA exchange functional alone, 
$\Ueff=\SI{2}{\electronvolt}$, \SI{2.25}{\electronvolt}, \SI{2.5}{\electronvolt} or
\SI{2.75}{\electronvolt} (see Fig.~\ref{fig:critical_temperarture_vs_vacancies}).
They all show an increase of the critical temperatures by increasing the oxygen 
content from $\delta= 0.3$ (\SI{10}{\atper}) towards stoichiometric SCO while 
their slopes varies qualitatively  between $\Ueff=\SI{2.25}{\electronvolt}$ and \sieV{2.5}.
The slope remained equal until $\Ueff=\SI{2.25}{\electronvolt}$ and even partially \sieV{2.5}
but became for larger $\Ueff$ reduced and matched well the trend of the experimental results.
We note that for $\Ueff=\SI{2.5}{\electronvolt}$ 
only the combined effect of $\Ueff$ and a larger amount of \vo is enough to obtain
the experimental trend, whereas $\Ueff=\SI{2.75}{\electronvolt}$ lies in the region of constant 
$\Tc^\text{MC}(\Ueff)$. For the latter, the influence of the oxygen vacancies alone 
to $\Tc^\text{MC}(\delta)$ was visible and it agreed well with the measurements
(see Fig.~\ref{fig:critical_temperature_vs_U}).} 

\add{However, these results were
just valid in a small range of oxygen-deficiency, otherwise \sco{3-\delta}
becomes unstable and forms different structure. This is also visible 
in the observed magnetic ground states depicted in Fig.~\ref{fig:critical_temperature_vs_U}.
At \SI{15}{\atper} oxygen-deficiency and $\Ueff\geq\SI{2.5}{\electronvolt}$, 
the resulting magnetic ground state was antiferromagnetic.}  This chemical
composition is already close to the ordered brownmillerite structure
\sco{2.5}, which is antiferromagnetic with a high N\'eel temperature
of $T_\text{N}=\SI{570}{\kelvin}$.\cite{Takeda1972} Although \add{the
cubic structure is} not the appropriate equilibrium structure at
this oxygen concentration, the \add{difference in} chemical composition might lead
already to an AFM order with \add{a higher} critical 
temperature. \add{Such a transition from a FM to an AFM ground state
can be caused in SCO e.g. by strain.\cite{Lee2011}  
A strain in general varies the unit cell volume and might therefore 
\add{modify} the magnetic coupling as well.}
\begin{figure}
 \includegraphics[width=\columnwidth]{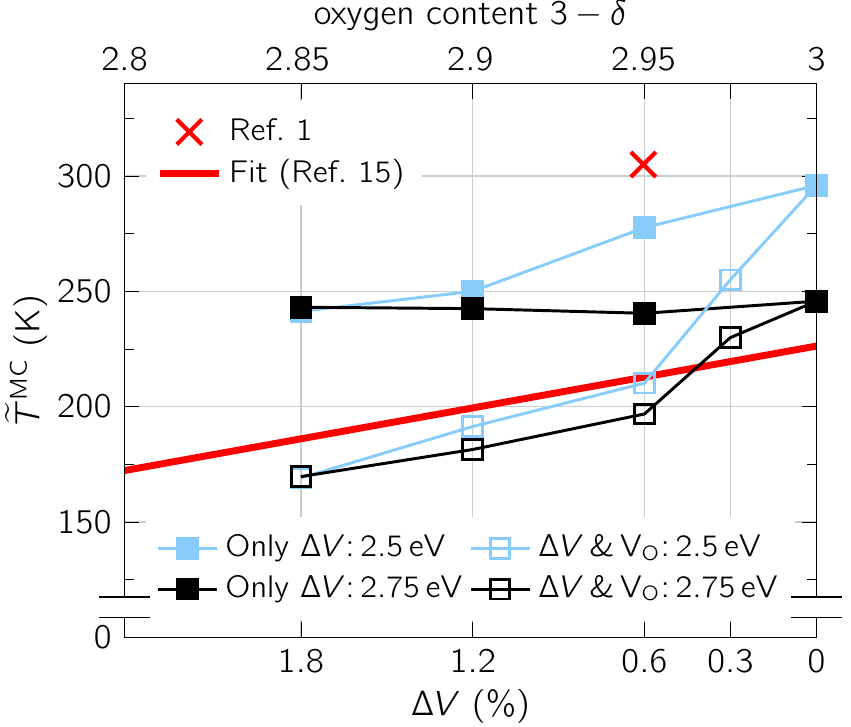}
 \caption{(Color online) \add{The theoretical critical temperature (DLM model, GGA$+U$) of
   \sco{3-\delta} in dependence of the volume change $\Delta V$ (filled squares; lower axis; 
   reversed to match other figures).
   The volume change is correlated with the oxygen-deficiency (upper axis)   
   via the experimental results of Ref.~\onlinecite{Balamurugan2006prb} (see text). 
   Both effects are also combined (open squares). 
   $\Ueff$ is \sieV{2.5} (blue) or \sieV{2.75} (black), respectively. 
   The magnetic ground state is always FiM.}}
   \label{fig:critical_temperarture_vs_vacancies_and_delta}
\end{figure} 

\subsection{Hydrostatic volume changes}
\label{sec:volume_effect}
\add{Up to now all calculations were performed with the fixed lattice
constant $a_\text{ref}$, but measured lattice constants indicate a volume expansion
in consequence of oxygen-deficiency (see examples in Fig.~\ref{fig:V vs delta exp}).} 
To estimate the influence of this volume enhancement to the 
critical temperature $\Tc^\text{MC}(\Delta V)$,
we choose from Fig.~\ref{fig:V vs delta exp} the largest volume expansion as reference.
\cite{Balamurugan2006prb}
So, we scaled the lattice parameters of the cubic unit cell for defect-free SCO
up to \SI{1.8}{\percent} and calculated $\Tc^\text{MC}$ in \rspm (see filled squares and the lower axis in
Fig.~\ref{fig:critical_temperarture_vs_vacancies_and_delta}).  
For $\Ueff=\sieV{2.5}$, the increasing distance between the Co atoms
reduces their magnetic interaction and the critical temperature, 
again similar as $\Ueff$ or \vo. Hence, $\Tc^\text{MC}(\Delta V)$
remains for $\Ueff=\sieV{2.75}$ almost constant.

\add{On the other hand, the volume expansion is correlated to a particular oxygen
content (see Fig.~\ref{fig:V vs delta exp}).
For simplicity, it is linearly interpolated from Ref.~\onlinecite{Balamurugan2006prb}
to derive in Fig.~\ref{fig:critical_temperarture_vs_vacancies_and_delta} the upper axis
(open squares). The combination of volume expansion and oxygen-deficiency leads for 
$\Ueff=\sieV{2.5}$ to a similar quantitative curve as for $\Tc^\text{MC}(\delta)$
but the observed kink appears already for $\delta=0.05$ or $\Delta V=\SI{0.6}{\percent}$,
respectively. In contrast, the qualitative and quantitative agreement of 
the variation of $\Tc^\text{MC}$ for $\Ueff=\sieV{2.75}$ is still in place.
Finally, we note that although the good agreement with Ref. \onlinecite{Balamurugan2006prb}
of the theoretically obtained $\Tc^\text{MC}$, the variation $\Delta V$ was the upper boundary 
of Fig.~\ref{fig:V vs delta exp}. 
For example, in Ref.~\onlinecite{LeToquin2006} the oxygen-deficiency of $\delta=0.16$
with respect to their value at $\delta=0$ is correlated to a much smaller volume expansion 
of $\Delta V= \SI{0.6}{\percent}$. As a result the slope of $\Tc^\text{MC}(\delta)$
changes drastically as well. }

\section{Conclusions}
We conclude that for cubic SrCoO$_{3-\delta}$ \add{the inclusion 
of temperature effects via the DLM model and a} small correlation
correction parameter of $\Ueff\approx\SI{2.75}{\electronvolt}$ is
\add{necessary} to describe the measured magnetic properties\add{, such as} the
magnetic moment or\add{, in particular,} the Curie temperature. 
Such \add{values of $U_\mathrm{eff}$ are} expected for a metallic system. 
\add{In the studied compound, we observe mainly the band magnetism,} 
which is reduced by oxygen-mediated superexchange.  
Furthermore, our calculations agree with the atomic
multiplet calculations and the picture of the intermediate spin state
of Potze \etal\cite{Potze1995}

\add{On the other side, oxygen vacancies can drastically alter
the magnetic properties.\cite{Balamurugan2006prb} 
The} simple means of the coherent potential approximation were
successful to model qualitatively \add{and quantitatively} 
the experimentally observed reduction
of $\Tc$ induced by oxygen vacancies.\cite{Balamurugan2006prb} 
They are one of the most
important types of defects in oxides -- even single crystalline
samples might be not completely stoichiometric.\cite{Long2011jpcm}
The randomly distributed vacancies weaken the Co-O-Co bonds \add{and} reduce
the exchange coupling\add{, similar to electronic correlations.} 
\add{The same} behavior was also observed for a possible increase of the unit
cell volume.

\section{Acknowledgments}
This work was funded by the \textit{Sonderforschungsbereich} SFB 762,
'Functionality of Oxide Interfaces'. We gratefully acknowledge
fruitful discussions with Igor V. Maznichenko and Alberto Marmodoro.

\bibliography{../../bib/jprb,../../bib/lib}
\bibliographystyle{apsrev4-1}
\end{document}